%% file: itwist18.tex
\documentclass[a4paper,10pt,twocolumn]{article}
\usepackage[english]{babel}
\usepackage[utf8]{inputenc}
\usepackage[T1]{fontenc}

\usepackage[top=1.5cm, left=1.5cm, right=1.5cm, bottom=1.5cm]{geometry}

\renewenvironment{abstract}{\bf\small {\em\ Abstract---}}{}

\usepackage{amsfonts,amssymb,amsmath,amsthm}
\usepackage{graphicx}
\usepackage[footnotesize]{caption}

\usepackage{amsfonts,amssymb,amsmath,amsthm}

\usepackage[utf8]{inputenc} % allow utf-8 input
\usepackage[T1]{fontenc}    % use 8-bit T1 fonts
\usepackage{hyperref}       % hyperlinks
\usepackage{url}            % simple URL typesetting
\usepackage{booktabs}       % professional-quality tables
\usepackage{amsfonts}       % blackboard math symbols
\usepackage{nicefrac}       % compact symbols for 1/2, etc.
\usepackage{microtype}      % microtypography

\usepackage{subcaption}
\usepackage{tikz,pgfplots}

\usepackage{scalefnt}
\newcommand{\eqdef}{\ensuremath{\stackrel{\mbox{\upshape\tiny def.}}{=}}}

\usepackage{amsmath}
\usepackage{amsthm}
\def\R{\mathbb{R}}
\def\B{\mathcal{B}}

\def\A{\mathcal{A}}
\def\C{\mathcal{C}}
\def\E{\mathcal{E}}
\def\F{\mathcal{F}}
\def\H{\mathcal{H}}
\def\L{\mathcal{L}}

\def\argmin{\mathop{\mathrm{argmin}}}

\def\Id{\mathrm{Id}}

\newtheorem{assumption}{Assumption}

\usepackage{algorithm}
\usepackage{algpseudocode}
\usepackage{xcolor}

\usepackage{dsfont}
\def\keywordname{{\bfseries Keywords}}%
\def\keywords#1{\par\addvspace\medskipamount{\rightskip=0pt plus1cm
\def\and{\ifhmode\unskip\nobreak\fi\ $\cdot$
}\noindent\keywordname\enspace\ignorespaces#1\par}}
\title{A scalable estimator of sets of integral operators}

\author{Valentin Debarnot$^1$, Paul Escande$^2$ and Pierre Weiss$^{1}$.\\
  \footnotesize $^1$ITAV, CNRS, Toulouse, France.\ $^2$John Hopkins University, Baltimore, USA.  } \date{\empty} % no need for a date

\begin{document}

\maketitle

\begin{abstract} 
We propose a scalable method to find a subspace $\widehat{\H}$ of low-rank tensors that simultaneously approximates a set of integral operators. The method can be seen as a generalization of the 
Tucker-2 decomposition model, which was never used in this context.
In addition, we propose to construct a convex set $\widehat \C \subset \widehat \H$ as the convex hull of the observed operators. It is a minimax optimal estimator under the Nikodym metric. 
We then provide an efficient algorithm to compute projection on $\widehat \C$. 
We observe a good empirical behavior of the method in simulations.
The main aim of this work is to improve the identifiability of complex linear operators in blind inverse problems.
\end{abstract}

% \keywords{Operator estimation, tensor decomposition, Tucker 2 decomposition, low-rank approximation, convex hull.}

\section{Introduction}\label{sec:intro}
\input{Introduction.tex}

\section{Operator representations}\label{sec:approx}
\input{Representation.tex}

\section{Subspace estimation}\label{sec:algo}
\input{Subspace.tex}

\section{Convex hull estimation} \label{sec:convex_hull}
\input{ConvexHull.tex}

% \section{Convergence rates} \label{sec:convex_hull}
% \input{Estimator.tex}

\subsubsection*{Acknowledgments}
This work was supported by the Fondation pour la Recherche Médicale (FRM grant number ECO20170637521).
The authors wish to thank Thomas Bonnafont for a preliminary study of the problem.
They wish to thank Thomas Mangeat for inspiring discussions.

% %% You can make the bibliography smaller
% \begin{thebibliography}{10}
% \bibitem{dummyref}
% A. Gebric and H. Armonic, 
% \newblock ``An introduction to humorous mathematics'',
% \newblock The Journal of Very Bad Jokes, {\bf 6}(4):123--134, 2006
% 
% \bibitem{lorem}
% C. Adams
% \newblock ``What does the filler text "lorem ipsum"
% mean?'', The Straight Dope, 2001
% \end{thebibliography}
\bibliographystyle{plain}
\bibliography{Bib}

\end{document}

%% file: Introduction.tex
In many measurement devices, a signal $u$ living in some Hilbert space $\B_n$ of dimension $n$ is probed indirectly using an operator $H:\B_n\to \B_m$, where $\B_m$ is a Hilbert space of dimension $m$\footnote{In all this paper, we assume that the operators are defined in finite dimensional spaces. An extension to infinite dimensional Hilbert spaces is feasible but it requires additional discretization procedures. We decided to skip this aspect to clarify the exposition.}. This yields a measurement vector $y\in \B_m$ defined by
\begin{equation}
 y = P(Hu),
\end{equation}
where $P$ is some perturbation of the measurements (e.g. additive noise, modulus for phase retrieval, quantization,...).
% Solving an inverse problem consists in recovering an approximation $\hat u$ of the signal $u$ using the measurements $y$.

 %\cite{vogel2002computational}
% When $H$ is known, many efficient solutions are now available. Unfortunately, in many cases, only a crude estimate of $H$ is available or it is completely unknown. This is the field of \emph{blind inverse problems}. In that case, finding a reasonable approximation is far more involved and despite significant progresses in the last fews years \cite{ahmed2014blind}, it can be still be considered as a mostly unresolved problem. 

In this paper, we are interested in a common situation where $H$ is unknown, but lives in a subset $\mathcal{C}$ of a low dimensional vector space $\H$. Our objective is to estimate both the set $\C$ 
and the vector space $\H$ from a sampling set of operators $(H_l)_{l \in L}$ in $\mathcal{C}$. The interest is that determining a low dimensional set of operators with a small volume can significantly 
ease the problem of operator identification in blind inverse problems. While our primary motivation lies in the field of inverse problems, this problem can also be understood as a generic problem of 
approximation theory.

\subsection{Contributions}

The simplest approach to find a low dimensional vector space of operators $\mathcal{H}$ is to apply the singular value decomposition (SVD) to a matrix formed by concatenating each vectorized operator 
$H_l$. However, this approach is infeasible in practice due to the huge dimension of matrix operators (up to dimension $10^{18}$ for 3D images of size $1024\times1024\times1024$). 

In this work, we therefore assume that the operators $H_l$ are given in a form suitable for computations. In essence, we will assume that they can be well approximated by low rank matrices or more 
generally tensors.

The principle of our approach is to find two orthogonal bases $(e_i)_{i\in I}$ and $(f_j)_{j \in J}$ such that all the operators  $H_l$  can be decomposed simultaneously as $H_l \simeq \sum_{i\in I, j 
\in J} \gamma_{i,j,l} e_i \otimes f_j$ for a well chosen matrix $\gamma_{:,:,l}\in \R^{|I| \times |J|}$. The estimate $\widehat{\mathcal{H}}$ of the vector space of operators $\mathcal{H}$ can then be 
identified with $E\otimes F \eqdef\mathrm{span}(e_i\otimes f_j, i\in I, j\in J)$. In addition, we propose to construct an estimate $\widehat{\mathcal{C}}$ of $\mathcal{C}$, as the convex hull (in a 
matrix space) of the coefficients $(\gamma_{:,:,l})_{l \in L}$. To make further use of this convex hull, we design a fast projection algorithm on $\widehat{\mathcal{C}}$.

Many of the tools used in this work are well established and its main originality is to bind them together to answer a question of high practical importance.

\section{Notation}

In all the paper, $I$, $J$, $K$ and $L$ are the sets of integers ranging from 1 to $|I|$, $|J|$, $|K|$ and $|L|$. 
We assume that $u\in \B_n$ is defined over a set $X$ of cardinality $n$ and that $Hu\in \B_m$ is defined over a set $Y$. % and $u(x)$ denote the value of $u$ at $x$.
% Similarly, we let $Y$ denote the space associated to $\B_m$. 
The letter $H$ can either refer to an operator $H:\B_n \rightarrow \B_m$ or its matrix representation in an arbitrary orthogonal basis. 
% The entries in the matrix representation will be denoted $H(y,x)$.
The Frobenius norm of $H$ is defined by $ \|H\|_F:= \sqrt{\mathrm{tr}(H^* H)}$. It is invariant by unitary transforms.
The tensor product between two vectors $a\in \B_m$ and $b\in \B_n$ is defined by $(a\otimes b)(y,x)=a(y)b(x)$.
The notation $\odot$ stands for the element-wise (Hadamard) product if $X$ has a group structure and $a_1,a_2\in \B_n$,  $a_1\star a_2$ denotes 
the convolution operator between $a_1$ and $a_2$.
We let $\Delta_N=\{x\in\R^{N},\; \sum_{i=1}^N x_i=1,x_i\geq 0\}$ denote the simplex of dimension $N$.

%% file: Representation.tex
A critical assumption in this work is that many operators can be approximated in a form tractable for computations. 
% The simplest assumption is to state that every operator $H_l$ is well approximated by a low-rank tensor of the form 
% $H_l= \sum_{k\in K} \alpha_{k,l} \otimes \beta_{k,l}$, for some vectors $\alpha_{k,l}\in \B_m$ and $\beta_{k,l}\in \B_n$, with $|K|\ll \min(m,n)$. Unfortunately, many observation operators met in 
% practice are concentrated along their diagonal, making this assumption unrealistic.
The most generic assumption on $H_l$ can be formulated as follows. Let $\L_1$ denote the set of operators from $\B_n$ to $\B_m$ and $\L_2$ denote another vector space.
\begin{assumption}
 There exists a left invertible linear mapping $T:\L_1 \to \L_2$ such that each sample $H_l \in \L_1$ satisfies:
 \begin{equation}
  S_l = T(H_l) = \sum_{k\in K} \alpha_{k,l}\otimes \beta_{k,l},
 \end{equation}
 where for all $l\in L$, the sets $(\alpha_{k,l})_{k} \in \A$ and $(\beta_{k,l})_{k} \in \B$, where $\A$ and $\B$ are subsets of $\B_m^{|K|}$ and 
$\B_n^{|K|}$ respectively.
\end{assumption}
This general formulation encompasses the usual low-rank assumption by taking $T=\Id$, but also product-convolution expansions \cite{escande2017approximation} that transform the usual 
kernel representation into the spatially varying impulse response.  
It also encompasses Hierarchical matrices \cite{bebendorf2008hierarchical} where $\A$ and $\B$ would incorporate multi-resolution support constraints.

%% file: Subspace.tex
The aim of this section is to provide an efficient and robust method to find a low dimensional basis of operators that allows to approximate 
simultaneously all the sampled representations $(S_l)_{l\in L}$.
Given two spaces of admissible bases $\E$ and $\F$ and two sets of integers $I$ and $J$, our aim is to solve the following problem:
\begin{equation}\label{eq:opt_algo}
\argmin_{\substack{(e_i)_{i\in I} \in \E \\ (f_j)_{j\in J} \in \F }} \frac{1}{2}\sum_{l\in L} \|\Pi_{E\otimes F}( S_l) - 
 S_l \|_{F}^2\eqdef \phi(E,F) ,
\end{equation}
where $\Pi_{E\otimes F}$ is the projection onto $E\otimes F$.
The sets $\E$ and $\F$ can encompass various structures according to applications, such as orthogonality or support constraints.

In the particular case where $\E$ and $\F$ are simply sets of orthogonal families of cardinality $|I|$ and $|J|$, Problem \eqref{eq:opt_algo} coincides 
exactly with the Tucker2 model. This problem is also linked with various other decompositions, we refer to the review paper \cite{kolda2009tensor} for more insight on tensor decompositions. A natural 
approach to solve such a problem is to use an alternating minimization algorithm. However, Problem \eqref{eq:opt_algo} is a complex 
nonconvex problem and such an algorithm is not ensured to converge to the global minimizer. The proposed procedure is summarized in Algorithm \ref{algo:iter_algo}.
\begin{algorithm}
\caption{Alternating Least Squares (ALS)}\label{algo:iter_algo}
\begin{algorithmic}[1]
\Statex \textbf{Approximatively solve:} Problem \eqref{eq:opt_algo}
\Statex \textbf{INPUT}: $(S_l)_{l\in L}$, subspace constraints $\E$ and $\F$.
\Statex \textbf{OUTPUT}: $\hat E\in \E$, $\hat F \in \F$.
\Procedure{}{}
\State Initialization: $k=0$, $F_0$.
\While{stopping criterion not satisfied}
\State $\displaystyle E_k = \argmin_{E\in \E} \phi(E,F_k).$
\State $\displaystyle F_{k+1} = \argmin_{F\in \F} \phi(E_k,F).$
\State $k=k+1$
\EndWhile
\EndProcedure
\end{algorithmic}
\end{algorithm}

Computing the two subproblems, when one family is fixed, can be rather challenging. However, we can take advantage of the particular structure of the operators $S_l$. For instance, when $\E$ 
and $\F$ only contain orthogonal constraints, as is the case for the Tucker 2 model, these subproblems can be solved computing the SVD of a matrix of size $n\times |I||L|$ or $m\times |J||L|$.

The starting point of Algorithm \ref{algo:iter_algo} is critical since the problem \eqref{eq:opt_algo} is non-convex. In this work, we use the High Order Singular 
Value Decomposition (HOSVD) \cite{de2000multilinear} which can be seen as a generalization of the SVD for tensors. 
% It is commonly used to initialize alternating algorithms.

In practice we observe that the alternation in Algorithm \ref{algo:iter_algo} may be pointless. Numerical experiments on diffusion operators show that 1 iteration of the algorithm yields a cost function which remains 
mostly unchanged when iterating. In Figure \ref{fig:approx}, we see that HOSVD and Algorithm \ref{algo:iter_algo} are indistinguishable. 
The curve called SVD refers to the error made by computing the SVD of the full vectorized operators. This method yields the best possible approximation, but it is only feasible with very 
low-dimensional operators, since it does not exploit a particular operator structure. The increase of computing time with respect to the dimensions is displayed in Figure \ref{fig:time}. 

\def\scal{0.5}
% \begin{figure}
% \centering
%      \includegraphics[scale=\scal]{Images/approxVSnbelemts.png} 
%      \caption{Relative approximation error against number of approximated elements kept.}\label{fig:approx}
% \end{figure}
% \begin{figure}
% \centering
%      \includegraphics[scale=\scal]{Images/complexityVSdim.png} 
%      \caption{Computation time against number of approximated elements kept.}\label{fig:approx}
% \end{figure}

% \begin{figure}
%     \centering
%     \begin{subfigure}[b]{\scal\textwidth}
%         \centering
%         \includegraphics[width=\textwidth]{Images/approxVSnbelemts.png}
%         \caption{Relative approximation error versus the dimension $|I|$ of each basis.}\label{fig:approx}
%     \end{subfigure}
%     \hfill
%     \begin{subfigure}[b]{\scal\textwidth}
%         \centering
%         \includegraphics[width=\textwidth]{Images/complexityVSdim.png}
%         \caption{Computation time versus dimension of the problem (the operators are of size $n\times n$).}\label{fig:time}
%     \end{subfigure}
%     \end{figure}

% \begin{figure}
% \centering
%      \includegraphics[scale=\scal]{Images/approxVSnbelemts.png} 
%      \caption{Relative approximation error against number of approximated elements kept.}\label{fig:approx}
% \end{figure}
% \begin{figure}
% \centering
%      \includegraphics[scale=\scal]{Images/complexityVSdim.png} 
%      \caption{Computation time against number of approximated elements kept.}\label{fig:approx}
% \end{figure}

\begin{figure}
    \centering
    \begin{subfigure}[b]{\scal\textwidth}
        \centering
\input{Images/approxVSbasis.tex}
        \caption{Relative approximation error versus the dimension $|I|$ of each basis.}\label{fig:approx}
    \end{subfigure}
    \hfill
    \begin{subfigure}[b]{\scal\textwidth}
        \centering
\input{Images/compVSdim.tex}
        \caption{Computation time in seconds versus dimension of the problem (the operators are of size $n\times n$).}\label{fig:time}
    \end{subfigure}
    \end{figure}

%% file: Images/approxVSbasis.tex
% This file was created by matlab2tikz.
%
%The latest updates can be retrieved from
%  http://www.mathworks.com/matlabcentral/fileexchange/22022-matlab2tikz-matlab2tikz
%where you can also make suggestions and rate matlab2tikz.
%
\definecolor{mycolor1}{rgb}{0.00000,0.44700,0.74100}%
\definecolor{mycolor2}{rgb}{0.85000,0.32500,0.09800}%
\definecolor{mycolor3}{rgb}{0.92900,0.69400,0.12500}%
\definecolor{mycolor4}{rgb}{0.49400,0.18400,0.55600}%
\begin{tikzpicture}

\begin{axis}[%
width=0.19*15.5in,
height=0.08*8.243in,
at={(2.6in,1.113in)},
scale only axis,
xmin=0,
xmax=30,
xtick={ 0,  5, 10, 15, 20, 25, 30},
ymode=log,
ymin=1e-05,
ymax=1,
yminorticks=true,
axis background/.style={fill=white},
legend style={nodes={scale=0.57, transform shape},at={(0.03,0.02637)}, anchor=south west, legend cell align=left, align=left, draw=white!15!black}
]
\addplot [color=mycolor1, line width=1.0pt]
  table[row sep=crcr]{%
0	1.0000000001\\
1	1.0000000001\\
2	0.999300409924924\\
3	0.324623912901074\\
4	0.317056556656104\\
5	0.184056077074032\\
6	0.172357929432923\\
7	0.114978864798558\\
8	0.107367070126007\\
9	0.077299591083339\\
10	0.0684568602301221\\
11	0.0520153176439841\\
12	0.049355674974863\\
13	0.0391226838597677\\
14	0.0375925493216542\\
15	0.0306138740172332\\
16	0.0298092111614939\\
17	0.0249157235367448\\
18	0.0244297921451223\\
19	0.0208452951943343\\
20	0.0206057489242057\\
21	0.0178663898555092\\
22	0.0176830770680993\\
23	0.0155329201580052\\
24	0.0154165818087025\\
25	0.013698162473156\\
26	0.0135870254399458\\
27	0.0121843183940669\\
28	0.0121050757820664\\
29	0.010941990644817\\
30	0.0108651340511639\\
};
\addlegendentry{DCT}

\addplot [color=mycolor2, line width=1.0pt]
  table[row sep=crcr]{%
0	1.0000000001\\
1	0.137327597450563\\
2	0.107450481051978\\
3	0.0882166707250221\\
4	0.0711073770757524\\
5	0.0553373775799929\\
6	0.039632618203945\\
7	0.0132250284056845\\
8	0.0113212595769495\\
9	0.00966048459599604\\
10	0.00852090207529959\\
11	0.0073348200429984\\
12	0.00620854428321056\\
13	0.00540382973726175\\
14	0.00481232490838702\\
15	0.004291814287771\\
16	0.00390558636010502\\
17	0.003508845760069\\
18	0.00308777141662171\\
19	0.00272796401645454\\
20	0.00240108067467753\\
21	0.00208062412383162\\
22	0.00177954909547275\\
23	0.00144233143357063\\
24	0.00123822316879248\\
25	0.00109557033699831\\
26	0.000949220869304672\\
27	0.000835615276388997\\
28	0.000730419691329136\\
29	0.000643772792125066\\
30	0.000554327358783975\\
};
\addlegendentry{SVD}

\addplot [color=mycolor3, line width=1.0pt]
  table[row sep=crcr]{%
0	1.0000000001\\
1	0.180805231775682\\
2	0.148414290468672\\
3	0.124879615366116\\
4	0.105711642700944\\
5	0.0900624532371871\\
6	0.0731854442498878\\
7	0.0546223264323054\\
8	0.0375007838138064\\
9	0.01629167083448\\
10	0.0141586153660902\\
11	0.0122574881653653\\
12	0.0101341826319782\\
13	0.00761225182437162\\
14	0.00612515015870453\\
15	0.00436151675098819\\
16	0.00334593812282215\\
17	0.00214053738733054\\
18	0.00183193858955012\\
19	0.00148903985157751\\
20	0.0011526131164046\\
21	0.000829498731838044\\
22	0.000621549916388315\\
23	0.000407589247690169\\
24	0.000289204558343602\\
25	0.00018540396625119\\
26	0.000145125702455432\\
27	0.000109292731314894\\
28	8.05994322754784e-05\\
29	5.54836678536833e-05\\
30	4.10145946451406e-05\\
};
\addlegendentry{HOSVD}

\addplot [color=mycolor4, line width=1.0pt]
  table[row sep=crcr]{%
0	1.0000000001\\
1	0.180800889786389\\
2	0.14833982913449\\
3	0.124314798280974\\
4	0.105280289792966\\
5	0.0896306772017904\\
6	0.0727151078524719\\
7	0.0540211933872239\\
8	0.0374032087792655\\
9	0.0162497920000084\\
10	0.0140853633921305\\
11	0.0120738531642925\\
12	0.00994184734124122\\
13	0.00748701447621128\\
14	0.00605555375490952\\
15	0.00429187528351386\\
16	0.00331988089172823\\
17	0.00212649383397942\\
18	0.0018039623856738\\
19	0.00147323575225542\\
20	0.0011432797908602\\
21	0.000819422958898843\\
22	0.000619751750931686\\
23	0.000402378611798344\\
24	0.000288007091513217\\
25	0.000184853843496375\\
26	0.000144961233159926\\
27	0.000109163173215381\\
28	8.05702840032529e-05\\
29	5.54645323120494e-05\\
30	4.10033889224757e-05\\
};
\addlegendentry{Tucker 2 via ALS}

\end{axis}
\end{tikzpicture}%

%% file: Images/compVSdim.tex
% This file was created by matlab2tikz.
%
%The latest updates can be retrieved from
%  http://www.mathworks.com/matlabcentral/fileexchange/22022-matlab2tikz-matlab2tikz
%where you can also make suggestions and rate matlab2tikz.
%
\definecolor{mycolor1}{rgb}{0.00000,0.44700,0.74100}%
\definecolor{mycolor2}{rgb}{0.85000,0.32500,0.09800}%
\definecolor{mycolor3}{rgb}{0.92900,0.69400,0.12500}%
\definecolor{mycolor4}{rgb}{0.49400,0.18400,0.55600}%
\begin{tikzpicture}

\begin{axis}[%
width=0.19*15.5in,
height=0.08*8.158in,
at={(2.6in,1.101in)},
scale only axis,
xmin=0,
xmax=4500,
xtick={ 512, 1024, 2048, 4096},
ymin=0,
ymax=45,
axis background/.style={fill=white},
axis x line*=bottom,
axis y line*=left,
legend style={nodes={scale=0.57, transform shape},at={(0.03,0.237)}, anchor=south west, legend cell align=left, align=left, draw=white!15!black},
/pgf/number format/.cd, 1000 sep={}
]
\addplot [color=mycolor1, line width=1.0pt]
  table[row sep=crcr]{%
32	0.119386\\
64	0.119827\\
128	0.127188\\
256	0.214746\\
512	0.270481\\
1024	0.395956\\
2048	0.511117\\
4096	0.788949\\
};
\addlegendentry{DCT}

\addplot [color=mycolor2, line width=1.0pt]
  table[row sep=crcr]{%
32	0.21133\\
64	0.0417\\
128	0.111393\\
256	0.184746\\
512	0.775906\\
1024	2.51192\\
2048	10.424755\\
4096	44.13122\\
};
\addlegendentry{SVD}

\addplot [color=mycolor3, line width=1.0pt]
  table[row sep=crcr]{%
32	0.145392\\
64	0.120318\\
128	0.118247\\
256	0.082894\\
512	0.173606\\
1024	0.15424\\
2048	0.14411\\
4096	0.334734\\
};
\addlegendentry{HOSVD}

\addplot [color=mycolor4, line width=1.0pt]
  table[row sep=crcr]{%
32	0.727526\\
64	0.978989\\
128	1.153491\\
256	1.431922\\
512	2.833845\\
1024	3.208082\\
2048	5.830008\\
4096	8.937945\\
};
\addlegendentry{Tucker 2 via ALS}

\end{axis}
\end{tikzpicture}%

%% file: ConvexHull.tex
Assume that we observe a finite number of points sampled with the uniform measure over a convex compact set $\C$. 
It is shown in \cite{brunel} that the convex hull $\widehat \C$ of those points is a minimax optimal estimator under the Nikodym metric. 
In the Hausdorff metric it is not known if the convex hull is still minimax optimal, but it only differs from the lower-bound by a logarithmic factor. 
%Using this estimate $\hat C$ allows to further reduce the dimensionality (and volume) of a set of operators.

This observation motivated us to develop a projection algorithm on $\widehat{\C} = \mathrm{conv}(\hat S_l, l \in L)$, where $\hat S_l$ is the projection of $S_l$ onto the subspace $\widehat \H = \hat 
E 
\otimes \hat F$.
Let $H$ be a linear operator and let $S=T(H)$. Then the projection of $S$ onto the convex hull $\widehat \C$ can be written as the following optimization problem:
 \begin{align}\label{eq:conv_hull1}
  \argmin_{\lambda\in\Delta_{|L|}} \frac{1}{2}\left\|\sum_{l\in L}\lambda_l\hat S_l-S\right\|_F^2 = \argmin_{\lambda\in\Delta_{|L|}} \frac{1}{2}\left\|M \lambda -S\right\|_F^2,
 \end{align}
where $M:\lambda \to \sum_{l\in L} \lambda_l \hat S_l$.
This problem can be solved efficiently using an accelerated proximal gradient descent algorithm. 
It is summarized in Algorithm \ref{algo:conv_hull}. The projection on the $(|L|-1)$-dimensional simplex can be computed in linear time.
\begin{algorithm}[H]
\caption{Projection onto convex hull of operators}\label{algo:conv_hull}
\begin{algorithmic}[1]
\Statex \textbf{INPUT}: A set $(\hat S_l)_{l\in L}$ in basis $E\otimes F$.
\Statex \textbf{OUTPUT}: Projection of $S$ onto $\widehat \C$.
\Procedure{}{}
\State an initial guess $\lambda_{0}\in \Delta_{|L|}$.
\For{$k=1,2,\hdots, k_{end}$}
\State $\tilde \lambda_{k}=\Pi_{\Delta_{L}}\left(M^* (M \lambda_{k}-S) \right)$
\State $\lambda_{k+1}= \tilde \lambda_{k} + \frac{k-1}{k+2} \left(\tilde \lambda_k - \tilde \lambda_{k-1}\right)$
\EndFor
\State Return $M\lambda_{k_{end}}$
\EndProcedure
\end{algorithmic}
\end{algorithm} 
Algorithm \ref{algo:conv_hull} ensures a convergence rate in $O(1/k^2)$. 
It can be made scalable by expressing the problem within the reduced basis $\widehat H$ of size $|I||J|$. We discard those technical aspects due to space limitation.